\documentclass[aps,nofootinbib,12pt]{revtex4}
\usepackage{amsmath}
\usepackage{amsfonts}
\usepackage{amssymb}
\usepackage{color}
\usepackage{graphicx}
\usepackage{mathrsfs}
\usepackage{slashbox}

\def\bc{\begin{center}}
\def\nno{\nonumber}
\def\ec{\end{center}}
\def\be{\begin{eqnarray}}
\def\ee{\end{eqnarray}}

\definecolor{dyellow}{rgb}{1.,0.8,.0}
\definecolor{myblue}{rgb}{.1,.1,.7}
\definecolor{dcyan}{rgb}{.0,.6,.6}
\definecolor{dmagenta}{rgb}{0.6,0.0,0.6}
\definecolor{brown}{rgb}{0.6,0.2,0.}
\definecolor{darkblue}{rgb}{.0,.0,0.5}
\definecolor{darkred}{rgb}{0.75,0.0,0.0}
\definecolor{orange}{rgb}{1.,.6,.0}
\definecolor{dorange}{rgb}{0.8,.4,.0}
\definecolor{darkgreen}{rgb}{0.0,0.6,0.0}
\definecolor{purple}{rgb}{.4,.0,.4}
\definecolor{lightgrey}{rgb}{0.7, 0.7, 0.7}
\definecolor{grey}{rgb}{0.4, 0.4, 0.4}

\def\al{\alpha}

\def\eps{\epsilon}

\def\Dl{\Delta}
\def\La{\Lambda}

\def\r{\partial}

\newcommand\rha{\rightarrow}

\def\lor{\langle O_{(1)}\rangle}
\def\lOr{\langle O_{(2)}\rangle}

\begin{document}
\title{Analytical Studies on Holographic Insulator/Superconductor Phase Transitions }

\author{Rong-Gen Cai$^{~1}$ }\email{cairg@itp.ac.cn}
\author{Huai-Fan Li$^{~1,2,3}$} \email{huaifan.li@stu.xjtu.edu.cn}
\author{Hai-Qing Zhang$^{~1}$} \email{hqzhang@itp.ac.cn}
 \address{$^{1}$Key Laboratory of Frontiers in Theoretical Physics,\\
Institute of Theoretical Physics, Chinese Academy of Sciences,\\
   P.O. Box 2735, Beijing 100190, China\\
  $^{2}$Department of Physics and Institute of Theoretical
   Physics,\\
Shanxi Datong University, Datong 037009, China \\
 $^{3}$Department of
Applied Physics, Xi' an Jiaotong University, \\ Xi' an 710049,
China}

\begin{abstract}
We investigate the analytical properties of the s-wave and p-wave
holographic insulator/superconductor phase transitions at zero
temperature. In the probe limit, we analytically calculate the
critical chemical potentials at which the insulator/superconductor
phase transition occurs.  Those resulting analytical values
perfectly match the previous numerical values. We also study the
relations between the condensation values and the chemical
potentials near the critical point. We find that the critical
exponent for condensation operator is $1/2$ for both models. The
linear relations between the charge density and the chemical
potential near the critical point are also deduced in this paper,
which are qualitatively consistent with the previous numerical
results.

\end{abstract}
\maketitle




%

\section{\bf Introduction }

The AdS/CFT correspondence~\cite{Maldacena:1997re} provides a
powerful theoretical method to understand the strongly coupled field
theories. Recently, it has been proposed that the AdS/CFT
correspondence can also  be used to describe superconductor phase
transitions~\cite{Gubser:2008px,Hartnoll:2008vx}. Since the
condensed matter physics deals with the systems at finite charge and
finite temperature, from the AdS/CFT correspondence the dual gravity
should be described by a charged black hole.

The phase transition studied in~\cite{Gubser:2008px,Hartnoll:2008vx}
is indeed a holographic superconductor/metal phase transition. The
simplest model for holographic superconductors can be constructed by
an Einstein-Maxwell theory coupled to a complex scalar field. In
particular, when the temperature of the black hole is below a
critical temperature, the black hole will become unstable to develop
a scalar hair near the horizon. And this scalar hair will break the
U(1) symmetry of the system. From the AdS/CFT correspondence, the
complex scalar field is dual to a charged operator in the boundary
field theory. And the breaking of the U(1) symmetry in gravity will
cause a global U(1) symmetry breaking in the dual boundary theory.
This induces a superfluid (superconductor) phase transition
\cite{Weinberg:1986cq}.

The holographic insulator/superconductor phase transition was first
studied in \cite{Nishioka:2009zj}. In particular, they used a
five-dimensional AdS soliton background \cite{Horowitz:1998ha}
coupled to a Maxwell and scalar field to model the holographic
insulator/superconductor phase transition at zero temperature. The
normal phase in the AdS soliton is dual to a confined gauge theory
with a mass gap which resembles an insulator phase
\cite{Witten:1998zw}. When the chemical potential is sufficiently
large, the AdS soliton becomes unstable to forming scalar hair which
is dual to a superconducting phase in the boundary field theory. The
holographic insulator/superconductor phase transition was also
studied in \cite{Horowitz:2010jq,Basu:2011yg,Brihaye:2011vk}.

In this paper, using the variational method for the Sturm-Liouville
eigenvalue problems \cite{Siopsis:2010uq}, we analytically studied
the s-wave and p-wave holographic insulator/superconductor phase
transitions in probe limit at zero temperature. We constructed the
s-wave model with the Einstein-Maxwell-scalar field theory in an AdS
soliton background. The order parameter for s-wave is the scalar
operator. On the other hand, we built the p-wave model with the
Einstein-Yang-Mills theory coupled to an AdS soliton background with
the order parameter to be the vector operator. The condensation of
the order parameter represents the onset of the superconducting. We
analytically calculated the critical chemical potentials in both
s-wave and p-wave models and found that they were in perfect
agreement with the previous numerical values in
\cite{Nishioka:2009zj, Akhavan:2010bf}. We also analytically
obtained the relations between the condensation values of the
operators and the chemical potentials near the critical point
$\mu_c$, we found that the general critical exponent $1/2$ would
always appear, {\it i.e.} $\langle O\rangle\propto \sqrt{\mu-\mu_c}$
which was qualitatively consistent with the former numerical results
\cite{Nishioka:2009zj, Akhavan:2010bf}. The linear relation of the
charge density $\rho$ and $(\mu-\mu_c)$ near the critical chemical
potential was obtained analytically which was also qualitatively
consistent with the previous numerical results. Other analytical
studies on holographic superconductors  can be found in
\cite{Gregory:2009fj,Pan:2009xa,Arean:2010zw,Zeng:2010zn,Li:2011xj,Chen:2011en}.

The paper is organized as follows. We studied the holographic s-wave
insulator/superconductor phase transition in Sec.\ref{sect:swave}.
In particular we calculated the critical chemical potential for
operators of various conformal dimensions. We also calculated the
relations of condensed values of operators and the charge density
with respect to $(\mu-\mu_c)$. The same procedure was also employed
in Sec.\ref{sect:pwave} for the p-wave case. At last, we draw our
conclusions in Sec.\ref{sect:con}.

\section{\bf S-wave holographic insulator/superconductor phase transition}
\label{sect:swave}

We construct the model of holographic insulator/superconductor phase
transition with the Einstein-Maxwell-scalar action in
five-dimensional spacetime:
 \be S=\int d^5x\sqrt{-g}(R+\frac{12}{L^2}-\frac1
 4F_{\mu\nu}F^{\mu\nu}-|\nabla_{\mu}\psi-iqA_{\mu}\psi|^2-m^2|\psi|^2).\ee
where $L$ is the radius of AdS space-time.

Following Ref.\cite{Nishioka:2009zj}, in the probe limit, we setup
this model in the AdS soliton background \cite{Horowitz:1998ha}:
 \be \label{metric} ds^2=L^2\frac{dr^2}{f(r)}+r^2(-dt^2+dx^2+dy^2)+f(r)d\chi^2.\ee
where, $f(r)=r^2-r_0^4/r^2$. In fact, this soliton solution can be
obtained from a five-dimensional AdS Schwarzschild black hole by
making use of two Wick rotations.  The asymptotical AdS space-time
approaches to a $R^{1,2}\times S^1$ topology near the boundary. And
the Scherk-Schwarz circle $\chi\sim\chi+\pi L/r_0$ is needed in
order to have a smooth geometry. The geometry looks like a cigar
whose tip is at $r=r_0$ if we extract the coordinates $(r,\chi)$.
Because of the compactified direction $\chi$, this background
provides a gravity description of a three-dimensional field theory
with a mass gap, which resembles an insulator in the condensed
matter physics. The temperature in this background is zero.

For simplicity, we will make ansatz of the matter fields as
$A_t=\phi(r),\quad \psi=\psi(r)$, which is consistent with the
equations of motions (EoMs):
  \be \label{eomr}\r_r^2\psi+(\frac{\r_rf}{f}+\frac3
  r)\r_r\psi+(-\frac{m^2}{f}+\frac{q^2\phi^2}{r^2f})\psi=0,\\
  \label{eomphi}\r_r^2\phi+(\frac{\r_rf}{f}+\frac1 r)\r_r\phi-\frac{2q^2\psi^2}{f}\phi=0.\ee
The boundary conditions near the infinity $r\rha\infty$ are:
 \be
 \psi&=&\psi^{(1)}r^{-2+\sqrt{4+m^2}}+\psi^{(2)}r^{-2-\sqrt{4+m^2}}+\cdots,\\
 \label{bcphi}
 \phi&=&\mu-\frac{\rho}{r^2}+\cdots.
 \ee
where, $\psi^{(i)}=\langle O_{(i)}\rangle,~i=1,2$, $O_{(i)}$ are the
corresponding dual operators of $\psi^{(i)}$ in the boundary field
theory. The conformal dimensions of the operators are
$\Dl_{\pm}=-2\pm\sqrt{4+m^2}$. There are two alternative
quantizations for the scalar field in AdS$_5$, {\it i.e.} the
operators $O_{(i)}$ are all normalizable \cite{Klebanov:1999tb},  if
 \be 0<\sqrt{4+m^2}<1\Rightarrow -4<m^2<-3.\ee
 In this paper, we will always set $m^2=-15/4$ except in subsection \eqref{sect:genm},  in order to compare
 with the numerical results in Ref.\cite{Nishioka:2009zj}. $\mu$ and $\rho$
 are the corresponding chemical potential and charge density in the
 boundary field theory.

The boundary conditions at the tip $r=r_0$ are:
 \be\label{bcpsi}
 \psi&=&a+b\log(r-r_0)+c(r-r_0)+\cdots,\\
 \label{neumann}
 \phi&=&A+B\log(r-r_0)+C(r-r_0)+\cdots.\ee
In order to take the Neumann-like boundary conditions, we impose
$b=B=0$ to render the physical quantities finite, see
Ref.\cite{Nishioka:2009zj}. It is worth noting that unlike in the
AdS black holes, the gauge field here is finite at $r=r_0$, {\it
i.e.} $A_t\neq0$.

Following Ref.\cite{Nishioka:2009zj}, we will scale $r_0=1,~q=1$ in
the subsequent calculations. Let $z=1/r$, the EoMs \eqref{eomr} and
\eqref{eomphi} become
 \be\label{eomz}
 \psi''+(\frac{f'}{f}-\frac1
 z)\psi'+\frac{(-m^2+\phi^2z^2)}{z^4f}\psi&=&0,\\
 \label{eomzphi}
 \phi''+(\frac{f'}{f}+\frac1 z)\phi'-\frac{2\psi^2}{z^4f}\phi&=&0.\ee
where $'$ denotes the derivative with respect to $z$.

\subsection{The critical chemical potential $\mu_c$}

From the numerical analysis in \cite{Nishioka:2009zj}, we can see
that when the chemical potential $\mu$ exceeds a critical chemical
potential $\mu_c$, the condensations of the operators will turn out.
This can be viewed as a superconductor (superfluid) phase. However,
when $\mu<\mu_c$, the scalar field is zero and this can be
interpreted as the insulator phase because this system has a mass
gap, which is due to the confinement in the (2+1)-dimensional gauge
theory via the Scherk-Schwarz compactificaiton. Therefore, the
critical chemical potential $\mu_c$ is the turning point of this
holographic insulator/superconductor phase transition.

 When
$\mu\leq\mu_c$, the scalar field in nearly zero {\it viz.}
$\psi\sim0$, therefore we can analytically
 solve  the gauge field equation \eqref{eomzphi}
 \be \phi''+(\frac{f'}{f}+\frac1 z)\phi'\approx0,\ee
 The general solution is
  \be \phi=C_2+C_1\log(\frac{1-z^2}{1+z^2}),\ee
 where $C_2$ and $C_1$ are the integration constants. The Neumann boundary condition \eqref{neumann} near
  $z=1$  imposes  $C_1=0$, so  $\phi(z)\equiv
 C_2=\mu$. This means $\phi(z)$ is a constant if $\psi(z)=0$, and from the boundary
 condition \eqref{bcphi}
 near $z=0$  we can get that $\rho=0$. This analytical results are consistent with
 the numerical results in Figure 2 in Ref.\cite{Nishioka:2009zj}, where $\rho=0$ when $\mu<\mu_c$.

 \subsubsection{Operators of dimension $\Dl=3/2$}
 Now, we take $m^2=-15/4$ like in \cite{Nishioka:2009zj}. For this $m^2$, the operators
 $O_{(1)}$ and $O_{(2)}$ are all normalizable as we have mentioned above. In this subsection,
 we will focus on the operator $O_{(1)}$ of conformal dimension
 $\Dl=3/2$. We will discuss the operator $O_{(2)}$ in the next
 subsection.

 When $\mu\rha\mu_c$, the scalar EoM \eqref{eomz} becomes
 \be \psi''+(\frac{f'}{f}-\frac1
 z)\psi'+\frac{1}{z^4f}(\frac{15}{4}+\mu^2z^2)\psi=0.\ee
To solve this equation, we can introduce a trial function $F(z)$ for
$\psi(z)$ near $z=0$ as~\cite{Siopsis:2010uq}
   \be \label{psiz0} \psi|_{z\rha0}\approx\langle O_{(1)}\rangle z^{3/2}F(z),\ee
The boundary condition for $F(z)$ is $F(0)=1,~F'(0)=0$.  In this
case, it is easy to deduce the EoM of $F(z)$ as
 \be
 F''+\frac{4z^3}{z^4-1}F'+\frac{-9z^4+4\mu^2z^2}{4z^2(1-z^4)}F=0.\ee
Multiply $T(z)$ to both sides of the above equation, where
 \be T(z)=z^4-1,\ee
 we have the EoM of $F(z)$
 \be \frac{d}{dz}[(z^4-1)F']+\frac{9}{4}z^2F-\mu^2F=0.\ee
 We can define the following parameters
 \be k=z^4-1,\quad P=-\frac94 z^2,\quad Q=-1.\ee
Thus, the minimum eigenvalues of $\mu^2$ can be obtained from taking
variations with the following functional \cite{Gelfand:1963}
 \be
\mu^2=\frac{\int_0^1dz(kF'^2+PF^2)}{\int_0^1dz~ QF^2}\ee
 In order to estimate the minimum value of $\mu^2$, we use the trial function, $F(z)=1-\al z^2$, where $\al$ is
a constant.  The minimum appears as
 \be \mu^2_{\rm
min}=\frac{3}{20}\times\frac{863\sqrt{230}-14950}{\sqrt{230}-414}\Rightarrow
\mu_{\rm min}\approx 0.837,\ee when
$\al={3}(35-2\sqrt{230})/61\approx 0.230$. Therefore,
$\mu_c=\mu_{\rm min}\approx 0.837$, which is in perfect
 agreement  with the numerical value $\mu_c\approx 0.84$ in Figure 2
 of Ref. \cite{Nishioka:2009zj}.

\subsubsection{Operators of general dimensions and $\Dl=5/2$}
\label{sect:genm}

The operator $O_{(2)}$ is normalizable when $m^2>m_{\rm BF}^2=-4$,
where $m_{\rm BF}^2$ is the Breitenlohner-Freedman (BF) bound of the
mass square of scalar field in the AdS space-time. In this
subsection, we will calculate the critical chemical potential
$\mu_c$ for the general case $-4<m^2<0$.

Following the the steps in the preceding subsection, we introduce a
trial function $F(z)$ into $\psi(z)$ near $z=0$,
  \be
\psi|_{z\rha0}\approx \lOr z^{2+\sqrt{4+m^2}}F(z),\ee
 The boundary conditions also impose $F(0)=1,~F'(0)=0$.
 It is easy to obtain the EoM of $F(z)$ as
 \be
 F''+\frac{1+2\sqrt{4+m^2}-z^4(5+2\sqrt{4+m^2})}{z-z^5}F'+\frac{(8+m^2+4\sqrt{4+m^2})z^2}{z^4-1}F+
 \frac{\mu^2}{1-z^4}F=0.\nno\\\ee
In this case, we multiply to both sides of the above equation with
$T(z)$:
 \be T(z)=z^{1+2\sqrt{4+m^2}}(z^4-1),\ee
Thus the EoM of $F(z)$  reduces to \be
\frac{d}{dz}[\underbrace{z^{1+2\sqrt{4+m^2}}(z^4-1)}_kF']+\underbrace{z^{3+2\sqrt{4+m^2}}(8+m^2+4\sqrt{4+m^2})}_{-P}F
\underbrace{-z^{1+2\sqrt{4+m^2}}}_Q\mu^2F=0.\nno\\\ee
 And the minimum eigenvalue of $\mu^2$ can be obtained by variating
 the following functional:
\be \mu^2&=&\frac{\int_0^1dz(kF'^2+PF^2)}{\int_0^1dz~QF^2}\nno\\
&=&\bigg[\left(15 \alpha ^2-32 \alpha +17\right) m^6+\left(517
   \alpha ^2-1160 \alpha +671\right) m^4+\left(4456
   \alpha ^2-9792 \alpha +5940\right) m^2\nno\\&&+720 \left(15
   \alpha ^2-32 \alpha +20\right)+\sqrt{m^2+4}
   \left((\alpha -1)^2 m^6+\left(107 \alpha ^2-238
   \alpha +133\right) m^4\right.\nno\\&&\left.+2 \left(776 \alpha ^2-1728
   \alpha +1035\right) m^2+360 \left(15 \alpha ^2-32
   \alpha +20\right)\right)\bigg]\bigg/\nno\\&&\bigg[\left(\sqrt{m^2+4} m^2+9 m^2+30 \sqrt{m^2+4}+60\right)
   \left(\alpha ^2 m^2-2 \alpha  m^2+m^2+3 \sqrt{m^2+4}
   \alpha ^2+6 \alpha ^2\right.\nno\\&&\left.-8 \sqrt{m^2+4} \alpha -14
   \alpha +5 \sqrt{m^2+4}+10\right)\bigg]\ee
In order to estimate it, we have set $F(z)=1-\al z^2$.

\begin{figure}[]
\centering
 (A.)\includegraphics[scale=0.55]{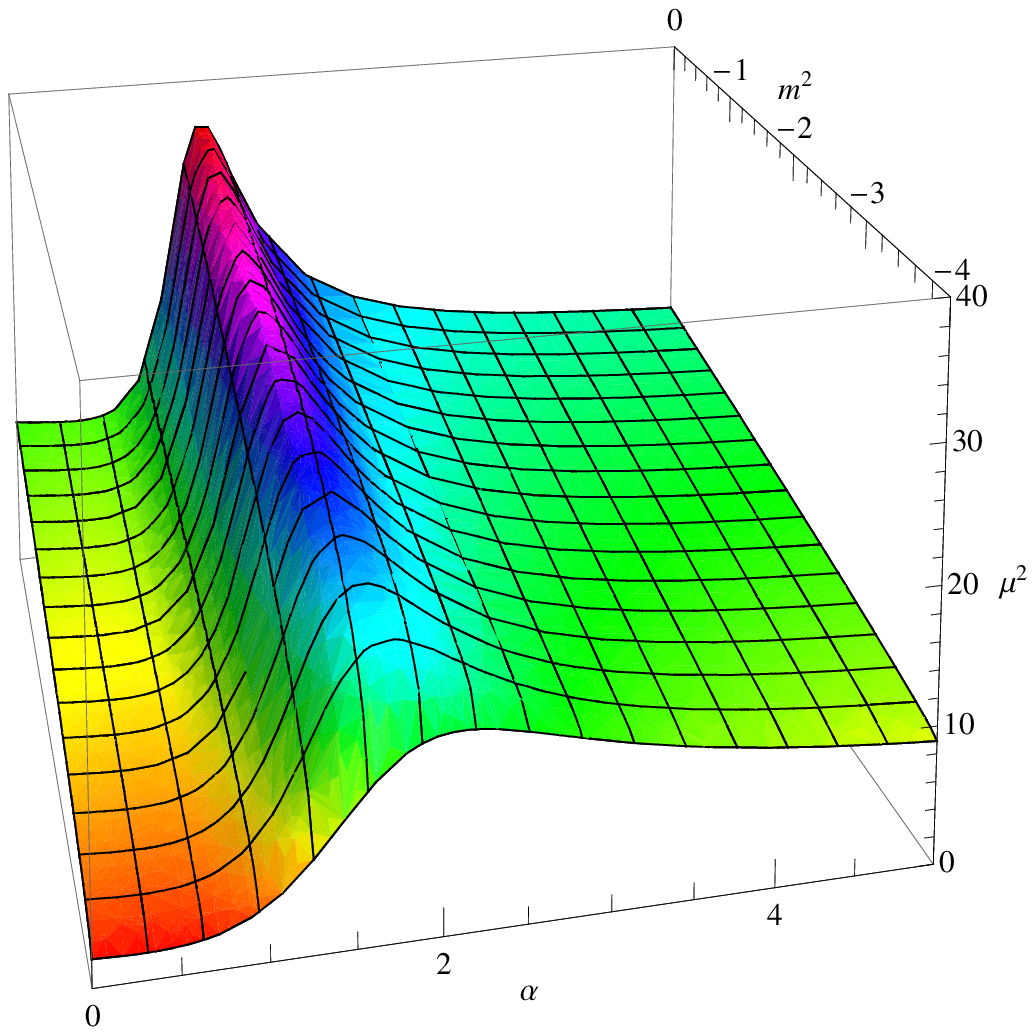}
(B.)\includegraphics[scale=0.75]{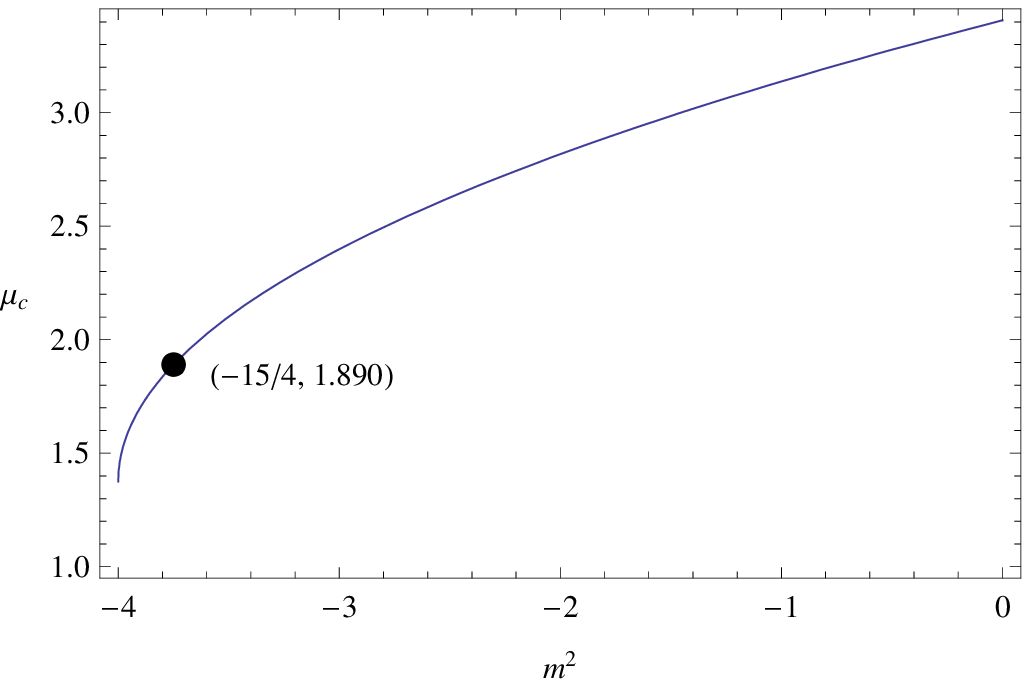}
 \caption{\label{musquare}(A.) The square of chemical potential as a function of
 $\al$ and $m^2$ where $-4<m^2<0$. (B.) The analytical relations between the critical
 chemical potential $\mu_c$ and the mass square of scalar field $m^2$. The black point
 represents the particular value for $m^2=-15/4$.}
\end{figure}

 We plot the function $\mu^2=\mu^2(\al, m^2)$ in the part A of
 Figure.\eqref{musquare}. It can be seen that there indeed exist the minimum values of
 $\mu^2$ and are very close to $\al\sim0$. In part B of
 Figure.\eqref{musquare}, we plot the relations of $\mu_c$ and
 $m^2$. It can be seen that when the mass square of the scalar field
 grows, the critical chemical potential grows too.

In particular, for $m^2=-15/4$ in Ref.\cite{Nishioka:2009zj}, we get
\be \mu^2_{\rm
min}=\frac{113190-1405\sqrt{462}}{15400+364\sqrt{462}}\Rightarrow
\mu_{\rm min}\approx 1.890,\ee when
$\al=5(63-2\sqrt{462})/303\approx 0.330$. This critical value
$\mu_c=\mu_{\rm min}\approx1.890$ is in well agreement with the
numerical values $\mu_c\approx 1.88$ in Figure 2 of
Ref.\cite{Nishioka:2009zj}. We have denoted this particular value in
part B of Figure.\eqref{musquare} as the black point.

\subsection{Relations of $\langle O\rangle$-$(\mu-\mu_c)$  and $\rho$-$(\mu-\mu_c)$ }

\subsubsection{Operators of dimension $\Dl=3/2$}

When $\mu$ is away from (but very close to) $\mu_c$, we can
substitute \eqref{psiz0} into the EoM of $\phi(z)$ \eqref{eomzphi}
as
 \be \phi''+(\frac{f'}{f}+\frac1
z)\phi'=\frac{2\lor^2F^2}{zf}\phi=0\ee
 Because near the critical chemical, the condensation of the
 operator is very small, we can expand $\phi(z)$ in $\lor$ as
  \be \phi(z)\sim
\mu_c+\lor\chi(z)+\cdots,\ee
 The boundary condition at the tip imposes the correction function $\chi(z)$ to be $\chi(1)=0$.
It is easy to deduce the EoM of  $\chi(z)$ as
 \be \label{chieom}\chi''-\frac{1+3z^4}{z-z^5}\chi'=\frac{2\lor \mu_c F^2}{zf}.\ee
Multiplying $T(z)$ to both sides of the above equation, where
 \be T(z)=\frac{z^4-1}{z}\ee
 The EoM of $\chi(z)$ is reduced to
 \be \frac{d}{dz}[\frac{z^4-1}{z}\chi']=-2\lor\mu_c F^2\ee
Making integration of both sides, we get
 \be\label{chilimit}
 \frac{z^4-1}{z}\chi'\bigg|_0^1=\frac{\chi'(z)}{z}\bigg|_{z\rha0}=-2\lor\mu_c\int_0^1dz
 (1-\al z^2)^2,\ee
 where we have used the trial function $F(z)=1-\al z^2$.\\
Near $z=0$, $\phi(z)$ can be expanded as
 \be\label{phiexp} \phi(z)\sim \mu-\rho z^2\approx
 \mu_c+\lor\bigg(\chi(0)+\chi'(0)z+\frac1 2\chi''(0)z^2+\cdots\bigg)\ee
Comparing  the coefficients of $z^0$ term in both sides of the above
formula, we get
 \be\label{muc} \mu-\mu_c\approx \lor\chi(0).\ee
 Besides, from the $z^1$ term in \eqref{phiexp}, we obtain that
 $\chi'(0)=0$. Therefore, from the EoM \eqref{chieom} and the
 boundary conditions of $\chi(z)$, we can solve $\chi(z)$ to be
  \be \chi(z)=&&\frac{\lor\mu_c}{60}\bigg(8 \alpha  \left(-\alpha  z^3+10 z+\alpha
   -10\right)+\pi  \left(3 \alpha ^2+10 \alpha
   +15\right)\nno\\&&-4 \left(3 \alpha ^2+10 \alpha
   +15\right) \arctan(z)+\left(3 \alpha ^2-10
   \alpha +15\right) \left(4 \log (z+1)-2 \log
   \left(z^2+1\right)\right)\nno\\&&+\left(-6 \alpha
   ^2+20 \alpha -30\right) \log2\bigg).\ee
 And then,
 \be \chi(0)=\frac{\lor\mu_c}{60}\bigg(8 (\alpha -10) \alpha +\pi  \left(3 \alpha ^2+10
   \alpha +15\right)+\left(-6 \alpha ^2+20 \alpha
   -30\right) \log2\bigg)\quad\quad.\ee
 Therefore, from \eqref{muc} we can deduce that
\be \mu-\mu_c\approx \frac{\lor^2\mu_c}{60}\bigg(8 (\alpha -10)
\alpha +\pi  \left(3 \alpha ^2+10
   \alpha +15\right)+\left(-6 \alpha ^2+20 \alpha
   -30\right) \log2\bigg)\quad\quad.\ee
 And further we have
   \be \label{conden}\lor\approx 1.940\sqrt{\mu-\mu_c},\ee
 when $\al=0.230$. This critical exponent $1/2$ for the condensation value
 and $(\mu-\mu_c)$ qualitatively match the numerical curves in
 Figure.1 of Ref.\cite{Nishioka:2009zj}.

Comparing the coefficients of the $z^2$ term in \eqref{phiexp}, we
get
 \be \rho=-\frac1 2\lor\chi''(0).\ee
From the EoM \eqref{chieom} and the formula \eqref{chilimit}, we can
obtain
 \be
 \chi''(0)=\frac{1+3z^4}{z-z^5}\chi'(z)\bigg|_{z\rha0}=\frac{\chi'(z)}{z}\bigg|_{z\rha0}=-2\lor\mu_c\int_0^1dz
 (1-\al z^2)^2.\ee
Thus, using the relation \eqref{conden}, it is easy to deduce that
 \be \rho\approx 2.700 (\mu-\mu_c),\ee
 when $\al=0.230$. This linear relation between the charge density
 $\rho$ and the chemical potential difference $(\mu-\mu_c)$
 qualitatively matches the numerical curves in Figure.2 of
 Ref.\cite{Nishioka:2009zj}. Moreover, this linear relation between $\rho$
 and $(\mu-\mu_c)$ can also be frequently seen in the numerical analysis
 in Ref.\cite{Horowitz:2010jq,Akhavan:2010bf}.

\subsubsection{Operators of dimension $\Dl=5/2$}

For the operators of dimension $\Dl=5/2$, we can follow the
preceding steps to get that
 \be\label{conden2} \lOr\approx1.801\sqrt{\mu-\mu_c}\ee
and
 \be\label{rhomu} \rho\approx 1.329 (\mu-\mu_c),\ee
 when $\al=0.330$. Eqs.~\eqref{conden2} and \eqref{rhomu} are
 qualitatively consistent with the numerical curves in Figure.1 and
 Figure.2 of Ref.\cite{Nishioka:2009zj}.

\section{P-wave holographic insulator/superconductor phase transition}
\label{sect:pwave}

Following \cite{Gubser:2008zu}, in this section we  consider a
five-dimensional SU(2) Einstein-Yang-Mills theory with a negative
cosmological constant. The action is
 \be S=\int d^5x
 \sqrt{-g}[\frac12(R-\La)-\frac14F^a_{\mu\nu}F^{a~\mu\nu}],\ee
where $F^a_{\mu\nu}$ is the field strength of the SU(2) gauge theory
and
$F^a_{\mu\nu}=\r_{\mu}A^a_{\nu}-\r_{\nu}A^a_{\mu}+\eps^{abc}A^b_{\mu}A^c_{\nu}$.
$a, b, c=(1,2,3)$ are the indices of the SU(2) Lie algebra
generator. $A^a_{\mu}$ are the components of the mixed-valued gauge
fields $A=A^a_{\mu}\tau^adx^{\mu}$, where $\tau^a$ are the
generators of the SU(2) Lie algebra with commutation relation
$[\tau^a, \tau^b]=\eps^{abc}\tau^c$. And $\eps^{abc}$ is a totally
antisymmetric tensor with $\eps^{123}=+1$.

As a consistent solution of the system, in the probe limit, the
background of the metric can also be an AdS soliton solution like
\eqref{metric} (We have scaled $L\equiv1, r_0\equiv1$),
 \be
ds^2= \frac{dr^2}{r^2g(r)}+r^2(-dt^2+dx^2+dy^2)+r^2g(r)d\chi^2,\ee
where we have set $f(r)=r^2g(r)=r^2(1-1/r^4)$.

We adopt the ansatz for the gauge field as~\cite{Gubser:2008wv}
  \be\label{ansatz} A(r)=\phi(r)\tau^3dt+\psi(r)\tau^1dx.\ee
 In this ansatz, the
gauge boson with nonzero component $\psi(r)$ along $x$-direction is
charged under $A^3_t=\phi(r)$. According to AdS/CFT dictionary,
$\phi(r)$ is dual to the chemical potential in the boundary field
theory while $\psi(r)$ is dual to the $x$-component of some charged
vector operator $O$. The condensation of $\psi(r)$ will
spontaneously break the U(1)$_3$ gauge symmetry and induce the
phenomena of superconducting on the boundary field theory.

Let $z=1/r$, the EoMs for $\phi(z)$ and $\psi(z)$ in the $z$
coordinate are
 \be
 \label{phieomp}\phi''+(\frac{g'}{g}-\frac1z)\phi'-\frac{\psi^2}{g}\phi=0,\\
 \label{psieomp}\psi''+(\frac{g'}{g}-\frac1z)\psi'+\frac{\phi^2}{g}\psi=0.\ee
The boundary conditions at infinity, {\it i.e.} $z\rha0$ are
 \be \phi&=&\mu-\rho z^2,\\
     \psi&=&\psi^{(0)}+\psi^{(2)}z^2,\ee
where, $\mu$ and $\rho$ can be interpreted as chemical potential and
charge density on the boundary field, respectively. $\psi^{(0)}$ and
$\psi^{(2)}$ represent the source and vacuum expectation value (vev)
of the dual operator on the boundary. We always set $\psi^{(0)}=0$
since we are interested in the case where the dual operator is not
sourced.

On the tip $z=1$, the boundary conditions are like the former ones
\eqref{bcpsi} and \eqref{neumann}:
 \be \psi&=&\al_0+\al_1(1-z)+\cdots,\\
    \phi& =&\beta_0+\beta_1(1-z)+\cdots.\ee

\subsection{The critical chemical potential $\mu_c$}

Following the analysis in the previous section,  when
$\mu\leq\mu_c$, $\psi$ is nearly zero, {\it i.e.} $\psi\sim0$.
Solving the equations \eqref{phieomp} we can get $\rho=0$ and
$\phi(z)={\rm Cons.}=\mu$ when $\mu<\mu_c$. This is consistent with
the numerical values in Figure.1 in Ref.\cite{Akhavan:2010bf}.

We can also introduce a trial function $F(z)$ into $\psi^{(2)}$ near
$z=0$,
 \be \psi|_{z\rha0}\sim \psi^{(2)}z^2\approx \lOr z^2F(z)\ee
 where we have set $\psi^{(2)}=\lOr$. And the boundary condition for
 $F(z)$ is $F(0)=1,~F'(0)=0$. Therefore, EoM of
 $F(z)$ is
 \be
 F''+\frac{3-7z^4}{z-z^5}F'-\frac{8z^2}{1-z^4}F+\frac{\mu^2}{g}F=0.\ee
Multiplying on both sides of the above equation with $T(z)$
 \be T(z)=z^7-z^3,\ee
 we have the EoM of $F(z)$ as
 \be \frac{d}{dz}[(z^7-z^3)F']+8z^5F-\mu^2z^3F=0.\ee
 We  define three parameters as follows
 \be k=z^7-z^3,\quad P=-8z^5,\quad Q=-z^3.\ee
 The minimum eigenvalues of $\mu^2$ can be obtained by varying the
 following functional
 \be \mu^2=\frac{\int_0^1dz(kF'^2+PF^2)}{\int_0^1dz~QF^2}.\ee
 It turns out that the minimum value is
 \be \mu^2_{\rm
 min}=\frac{16}{5}\times\frac{10-15\al+8\al^2}{6-8\al+3\al^2}\Rightarrow
 \mu_{\rm min}\approx 2.267 \ee
 when $\al=(18-\sqrt{134})/19\approx 0.338$. The critical value $\mu_c=\mu_{\rm min}\approx
 2.267$ is in great agreement with the numerical values $\mu_c\approx
 2.26$ in Figure.1 of Ref.\cite{Akhavan:2010bf}.

\subsection{Relations of $\langle O\rangle$-$(\mu-\mu_c)$  and $\rho$-$(\mu-\mu_c)$ }

 When $\mu\rha\mu_c$, the condensation value of $\psi(z)$ is very
 small, we can expand $\phi(z)$ in $\lOr$ as
 \be\label{phic} \phi\sim\mu_c+\lOr\chi(z)+\cdots.\ee
 where the boundary condition imposes $\chi(1)=0$.
Substituting \eqref{phic} into \eqref{phieomp}, we get the EoM of
$\chi(z)$
 as
 \be\label{chieomp}
 \chi''-\frac{1+3z^4}{z-z^5}\chi'-\lOr\mu_c\frac{z^4}{1-z^4}F^2=0.\ee
 Near $z=0$, we can expand $\phi(z)$ as
 \be\label{phiz} \phi=\mu-\rho z^2\approx
 \mu_c+\lOr(\chi(0)+\chi'(0)z+\frac12\chi''(0)z^2+\cdots).\ee
Comparing the coefficients of the $z^0$ term, we obtain
 \be\label{mucp} \mu=\mu_c+\lOr\chi(0).\ee
 And from the $z^1$ term, we know that $\chi'(0)=0$. We can solve
 $\chi(z)$ via the EoM \eqref{chieomp} and the boundary conditions of
 $\chi(z)$ to be
\be  \chi(z)=-\frac{\lOr\mu_c}{48}\bigg[\left(z^6+3 z^2-4\right)
\alpha ^2+\left(-4 z^4+8
   \log \left(\frac{2}{z^2+1}\right)+4\right)
   \alpha +6 \left(z^2-1\right)\bigg],\nno\\\ee
which gives
 \be
\chi(0)=\frac{\lOr\mu_c}{24}\bigg(3+2\al^2-2\al(1+2\log2)\bigg).\ee
Further we deduce from \eqref{mucp} that
 \be
\lOr\approx2.560\sqrt{\mu-\mu_c}=3.855\sqrt{\frac{\mu}{\mu_c}-1},\ee
when $\al=0.338$. Once again, this critical exponent $1/2$ is
qualitatively consistent with the numerical curves in Figure.1 of
Ref.\cite{Akhavan:2010bf}.

From the equation \eqref{chieomp}, we can have
 \be\label{chipp}
\chi''(0)=\frac{1+3z^4}{z-z^5}\chi'(z)\bigg|_{z\rha0}=\frac{\chi'(z)}{z}\bigg|_{z\rha0}.\ee
Multiplying $T(z)$ on both sides of \eqref{chieomp}, where
 \be
T(z)=\frac{z^4-1}{z}\ee we have
 \be \frac{d}{dz}[\frac{z^4-1}{z}\chi']=-\lOr\mu_cz^3F^2,\ee
which gives us with
 \be
 \frac{z^4-1}{z}\chi'(z)\bigg|_0^1=\frac{\chi'(z)}{z}\bigg|_{z\rha0}=-\lOr\mu_c\int_0^1dz~z^3F^2.\ee
From \eqref{chipp} we can obtain with $F(z)=1-\al z^2$ that
 \be \chi''(0)=-\lOr\mu_c(\frac1 4-\frac{\al}{3}+\frac{\al^2}{8}).\ee
Comparing the coefficients of the $z^2$ term of \eqref{phiz}, we
reach
 \be \rho=\frac12\lOr^2\mu_c(\frac1
 4-\frac{\al}{3}+\frac{\al^2}{8})\approx 1.126(\mu-\mu_c),\ee
 when $\al=0.338$. This linear relation between $\rho$ and
 $(\mu-\mu_c)$ is qualitatively consistent with the numerical curves
 in Figure.1 of Ref.\cite{Akhavan:2010bf}.

\section{Conclusions}
\label{sect:con}

In this paper, we have studied the analytical properties of the
s-wave and p-wave holographic insulator/superconductor phase
transitions at zero temperature. In particular, we have ignored the
back-reaction of the gauge field and the scalar field to the AdS
soliton background. When the chemical potential $\mu$ is lower than
the critical chemical potential $\mu_c$, the AdS soliton background
is stable and the dual field theory can be interpreted as an
insulator; However, when $\mu>\mu_c$ the AdS soliton background will
be unstable to form the condensations of the scalar field (in
s-wave) or the vector field (in p-wave). And the dual field theory
is in a superconducting phase.

 Actually, we have analytically
obtained the critical chemical potentials in both s-wave and p-wave
models. We found that the critical chemical potential $\mu_c$ we
obtained are perfectly in agreement with the previous numerical
values. In particular, for the scalar operator of conformal
dimension ${2+\sqrt{4+m^2}}$, we found that when $m^2$ grows, the
critical chemical potential $\mu_c$ grows as well. Besides, we
calculated the relations between the condensation values of the dual
operator and the chemical potential near $\mu_c$, and found that the
critical exponent of condensation operator is always $1/2$ in both
models, {\it i.e.} $\langle O\rangle\propto\sqrt{\mu-\mu_c}$. In
addition, we also obtained the linear relations between the charge
density and the chemical potential near $\mu_c$, which is also
qualitatively consistent with the previous numerical results. Our
results combining with others in the literature
\cite{Siopsis:2010uq,Zeng:2010zn,Li:2011xj} show that the analytic
method is quite powerful to study the holographic superconducting
phase transition near the critical point.

\acknowledgments
HFL and HQZ would like to thank Bin Hu, Zhang-Yu Nie and Yun-Long
Zhang for their helpful discussions and comments. HFL would be very
grateful for the hospitalities of the members in the Institute of
Theoretical Physics, Chinese Academy of Sciences. This work was
supported in part by the National Natural Science Foundation of
China (No. 10821504, No. 10975168, No.11035008 and No.11075098), and
in part by the Ministry of Science and Technology of China under
Grant No. 2010CB833004.


\end{document}